\documentclass{article}

\usepackage{amsmath,amssymb}

\usepackage[utf8x]{inputenc}

\usepackage{textcomp,marvosym}

\usepackage{cite}

\usepackage[table]{xcolor}

\usepackage{array}
\usepackage{graphicx}
\usepackage{changepage}

\usepackage[aboveskip=1pt,labelfont=bf,labelsep=period,justification=raggedright,singlelinecheck=off]{caption}

\date{}

\usepackage{booktabs}
\usepackage{enumerate} 
\usepackage{listings}
\usepackage{array}
\usepackage{hanging}
\usepackage{xcolor}
\usepackage{hyperref}

\begin{document}

\title{Anatomy of an online misinformation network}

\author{
Chengcheng Shao\textsuperscript{1,2*\Yinyang}
\and
Pik-Mai Hui\textsuperscript{2\Yinyang}
\and
Lei Wang\textsuperscript{2}
\and
Xinwen Jiang\textsuperscript{3}
\and
Alessandro Flammini\textsuperscript{2\ddag}
\and
Filippo Menczer\textsuperscript{2\ddag} 
\and
Giovanni Luca Ciampaglia\textsuperscript{4\ddag}
\\
\\
\textbf{1} College of Computer,\\ National University of Defense Technology, China
\\
\textbf{2} School of Informatics, Computing, and Engineering,\\ Indiana University, Bloomington, USA
\\
\textbf{3} The MOE Key Laboratory of Intelligent Computing\\ and Information Processing, Xiangtan University, China
\\
\textbf{4} Indiana University Network Science Institute,\\ Bloomington, USA
}


\maketitle

%
%
\centerline{\Yinyang{} These authors contributed equally to this work.}

\centerline{\ddag{} These authors also contributed equally to this work.}
\centerline{* Corresponding author. Email: sccotte@gmail.com 
}

\begin{abstract}
Massive amounts of fake news and conspiratorial content have spread over social media before and after the 2016 US Presidential Elections despite intense fact-checking efforts. 
How do the spread of misinformation and fact-checking compete?
What are the structural and dynamic characteristics of the core of the misinformation diffusion network, and who are its main purveyors? How to reduce the overall amount of misinformation? 
To explore these questions we built Hoaxy, an open platform that enables large-scale, systematic studies of how misinformation and fact-checking spread and compete on Twitter.
Hoaxy filters public tweets that include links to unverified claims or fact-checking articles. 
We perform $k$-core decomposition on a diffusion network obtained from two million retweets produced by several hundred thousand accounts over the six months before the election. 
As we move from the periphery to the core of the network, fact-checking nearly disappears, while social bots proliferate.
The number of users in the main core reaches equilibrium around the time of the election, with limited churn and increasingly dense connections. 
We conclude by quantifying how effectively the network can be disrupted by penalizing the most central nodes.  
These findings provide a first look at the anatomy of a massive online misinformation diffusion network.
\end{abstract}

\section{Introduction}

The viral spread of online misinformation is emerging as a major threat to the free exchange of opinions, and consequently to democracy. Recent Pew Research Center surveys found that 63\% of Americans do not trust the news coming from social media, even though an increasing majority of respondents uses social media to get the news on a regular basis (67\% in 2017, up from 62\% in 2016). Even more disturbing, 64\% of Americans say that fake news have left them with a great deal of confusion about current events, and 23\% also admit to passing on fake news stories to their social media contacts, either intentionally or unintentionally~\cite{Pew2016,Pew2017a,Pew2017b}.

Misinformation is an instance of the broader issue of abuse of social media platforms, which has received a lot of attention in the recent literature~\cite{Ratkiewicz:2011:TMS:1963192.1963301, W2015HiddenHands,ICWSM112850,Sampson:2016:LIS:2983323.2983697,Wu2016,Declerck2016,Kumar:2016:DWI:2872427.2883085,Varol2017,botornot_icwsm17,Ferrara:2016:RSB:2963119.2818717,FM8005,2017arXiv170707592S}. The traditional method to cope with misinformation is to fact-check claims. Even though some are pessimistic about the effectiveness of fact-checking, the evidence is still conflicting on the issue~\cite{ECKER2017185,Nyhan2016}. In experimental settings, perceived social presence reduces the propensity to fact-check~\cite{Jun06062017}. An open question is whether this finding translates to the online setting, which would affect the competition between low-and high-quality information. This question is especially pressing. Even though algorithmic recommendation may promote quality under certain conditions~\cite{2017arXiv170700574N}, models and empirical data show that low-quality information may be as likely to go viral as high-quality information in online social networks~\cite{Qiu2017,2017arXiv170707592S}.  

Technology platforms, journalists, fact checkers, and policymakers are debating how to combat the threat of misinformation~\cite{Wardle2016}. A number of systems, tools, and datasets have been proposed to support research efforts about misinformation. Mitra and Gilbert, for example, proposed CREDBANK, a dataset of tweets with associated credibility annotations~\cite{ICWSM1510582}. Hassan \emph{et al.}~\cite{Hassan:2015:DCF:2806416.2806652} built a corpus of political statements worthy of fact-checking using a machine learning approach. Some systems let users visualize the spread of rumors online. The most notable are TwitterTrails~\cite{Metaxas:2015:UTI:2685553.2702691} and RumorLens~\cite{ICWSM1510592}. These systems, however, lack monitoring capabilities. The Emergent site~\cite{Emergent.info} detected unverified claims on the Web, tracking whether they were subsequently verified, and how much they were shared. The approach was based on manual curation, and thus did not scale.

The development of effective countermeasures requires an accurate understanding of the problem, as well as an assessment of its magnitude~\cite{Ciampaglia2017,Lazer2017}. To date, the debate on these issues has been informed by limited evidence. 
Online social network data provides a way to investigate how human behaviors, and in particular  patterns of social interaction, are influenced by newsworthy events~\cite{Lu2014Network}. 
Studies of news consumption on Facebook reveal that users tend to confine their attention on a limited set of pages~\cite{del2016spreading,Schmidt21032017}.
Starbird demonstrates how alternative
news sites propagate and shape narratives around mass-shooting events~\cite{starbird2017examining}.

Articles in the press have been among the earliest reports to raise the issue of fake news~\cite{Silverman2016}. Many of these analyses, however, are hampered by the quality of available data --- subjective, anecdotal, or narrow in scope. In comparison, the internal investigations conducted by the platforms themselves appear to be based on comprehensive disaggregated datasets~\cite{FB-INFO-OPS,FB-LOW-Q}, but lack transparency, owing to the two-fold risk of jeopardizing the privacy of users and of disclosing internal information that could be potentially exploited for malicious purposes~\cite{TwitterBlog}. 

Motivated by these limitations, in previous work we presented a prototype of Hoaxy, an open platform for the study of the diffusion of misinformation and its competition with fact-checking~\cite{shao2016hoaxy}.
Here we build upon this prior effort, contributing to the debate on how to combat digital misinformation in two ways:

\begin{itemize}
    \item We describe the implementation and deployment of the Hoaxy system, which was first introduced in a 2016 demo~\cite{shao2016hoaxy}. The system has been collecting data on the spread of misinformation and fact checking from the public Twitter stream since June of 2016. It is now publicly available (\url{hoaxy.iuni.iu.edu}). Users can query the tool to search instances of claims and relative fact checking about any topic and visualize how these two types of content spread on Twitter. 
    \item We leverage the data collected by Hoaxy to analyze the diffusion of claims and fact-checks on Twitter in the run up to and wake of the 2016 US Presidential Election. This analysis provides a first characterization of the anatomy of a large-scale online misinformation diffusion network.
\end{itemize}

When studying misinformation, the first challenge is to assess the truthfulness of a claim. This presents several difficulties. The most important is scalability: it is impossible to manually evaluate a very large number of claims, even for professional fact-checking organizations with dedicated staff. Here we mitigate these issues by relying on a list of sources compiled by trusted third-party organizations. In the run-up to and wake of the 2016 US Presidential Elections, several reputable media and fact checking organizations have compiled lists of popular sources that routinely publish unverified content such as hoaxes, conspiracy theories, fabricated news, click bait, and biased, misleading content. In the remainder of the paper we informally refer to this content as ``claims.'' We manually assess that the great majority of the claims published by these sources, considered here, contain some form of misinformation or cannot be verified (see Methods).

Hoaxy retrieves the full and comprehensive set of tweets that share (i.e., include a link to) claims and fact-checks. These tweets are important because, by tracking them, we can observe how a particular piece of content spreads over the social network. It is important to note that Hoaxy collects 100\% of these tweets, not a sample. This lets us obtain, for any given piece of misinformation in our corpus, the full picture of how it spreads and competes with subsequent fact-checking, if any. 

In this paper we address three research questions:

\begin{itemize}
    \item{\textbf{RQ1:}} How do the spread of misinformation and fact-checking compete?
    \item{\textbf{RQ2:}} What are the structural and dynamic characteristics of the core of the misinformation diffusion network, and who are its main purveyors?
    \item{\textbf{RQ3:}} How to reduce the overall amount of misinformation?
\end{itemize}

We pose our first question (RQ1) to investigate whether those who are responsible for spreading claims are also exposed to corrections of those claims. Regretfully, only 5.8\% of the tweets in our dataset share links to fact-checking content --- a 1:17 ratio with misinformation tweets. We analyze the diffusion network in the run up to the election, and find a strong core-periphery structure. Fact-checking almost disappears as we move closer to the inner core of the network, but surprisingly we find that \emph{some} fact-checking content is being shared even inside the main core. Unfortunately, we discover that these instances are not associated with interest in accurate information. Rather, links to Snopes or Politifact are shared either to mock said publications, or to mislead other users (e.g., by falsely claiming that the fact-checkers found a claim to be true). This finding is consistent with surveys on the trust on fact-checking organizations, which find strong polarization of opinions~\cite{Brandtzaeg:2017:TDO:3134526.3122803}. 

Our second question (RQ2) is about characterizing the core of the claim diffusion network. We find the main core to grow in size initially and then become stable in both size and membership, while its density continues to increase. We analyze the accounts in the core of the network to identify those users who play an important role in the diffusion of misinformation. The use of Botometer, a state-of-the-art social bot detection tool~\cite{botornot_icwsm17}, reveals a higher presence of social bots in the main core. We also consider a host of centrality measures (in-strength, out-strength, betweenness, and PageRank) to characterize and rank the accounts that belong in the main core. Each metric emphasizes different subsets of core users, but interestingly the most central nodes according to different metrics are found to be similar in their partisan slant.

Our last question (RQ3) addresses possible countermeasures. Specifically we ask what actions platforms could take to reduce the overall exposure to misinformation. Platforms have already taken some steps this direction, by prioritizing high-quality over low-quality content~\cite{FB-LOW-Q,Google-LOW-Q}. Here we take a further step in this direction and investigate whether penalizing the main purveyors of misinformation, as identified by RQ2, yields an effective mitigation strategy. We find that a simple greedy solution would reduce the overall amount of misinformation significantly. 

All the analyses presented in this paper can be replicated by collecting data through the Hoaxy API~\cite{HoaxyAPIDoc} or downloading the network dataset at \url{doi.org/10.5281/zenodo.1154840}.

\section{Methods and data}

\subsection{Network core analysis}

The \emph{$k$-core} of a graph is formally defined as the maximal subgraph with nodes of at least degree $k$. In practice, $k$-core \emph{decomposition} uses a recursive procedure, given the $k$-core, for extracting the $(k+1)$-core by recursively removing all nodes with degree $k$. The nodes that have been removed constitute the \emph{$k$-shell}. The $k$-core decomposition is the sequence of $k$-cores of increasing values of $k$. Finally, the non-empty graph with maximum value of $k$ is called the \emph{main core}. Prior work has used $k$-core decomposition to probe the structure of complex networks~\cite{PhysRevLett.96.040601,Alvarez-Hamelin2008}. In the case of social networks, $k$-cores can be used to identify influential users~\cite{Kitsak2010}, and to characterize the efficiency of information spreading~\cite{conover12partisan}.

\subsection{Bot detection}

Social bots play an important role in the spread of misinformation~\cite{2017arXiv170707592S}. Researchers have built supervised learning tools to detect such automated accounts with high accuracy. We leverage such a tool, called Botometer~\cite{botornot_icwsm17}, to evaluate Twitter accounts. 

Botometer performs classification over a large set of features that include temporal, network, language, and sentiment signals. The classifier is trained in supervised fashion from a set of labeled examples. The set includes examples discovered with a honeypot and by human raters. Two classifiers are available, a standard one, which includes English-based language features, and a `universal' one, which does not include language features, and is thus applicable beyond English-speaking contexts. We use the standard classifier through a public API~\cite{Davis16BotOrNot,BotometerAPI}. 

\subsection{Claim verification}

Our analysis considers content published by a set of websites flagged as sources of misinformation by third-party journalistic and fact-checking organizations. We merged several lists of `misinformation' sources compiled by such organizations. It should be noted that these lists were compiled independently of each other, and as a result they have uneven coverage. However, there is some overlap between them. The full list is available online~\cite{HoaxyFAQ}. 

\begin{figure}
    \centering
    \includegraphics[width=0.4\columnwidth]{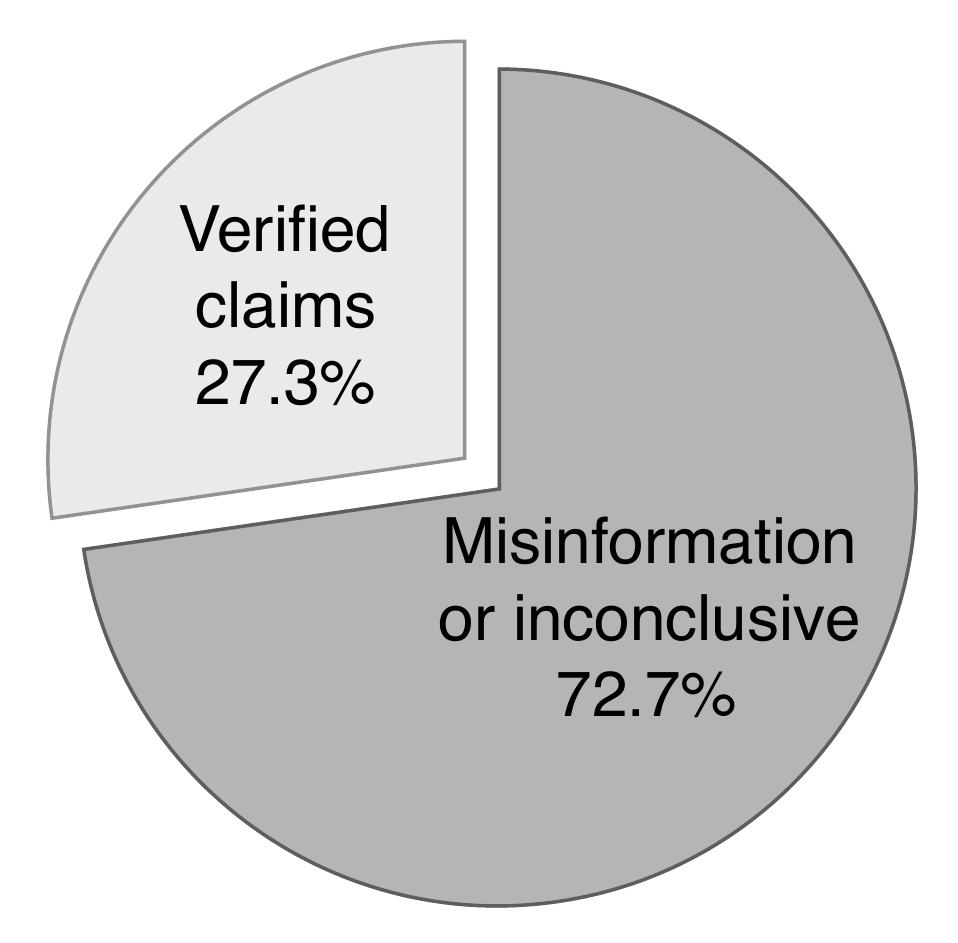}
    \caption{Verification based on a sample of 50 claims. We excluded six articles with no factual claim. Articles that could not be verified are grouped with misinformation.}
    \label{fig:piechart}
\end{figure}

The source-based approach relies on the assumption that most of the claims published by our compilation of sources are some type of misinformation, as we cannot fact-check each individual claim. 
To validate this assumption, we manually verified a random sample of 50 articles drawn from our corpus, considering only those sources whose articles were tweeted at least once in the period of interest. Each article was evaluated independently by two reviewers, with ties broken by a third reviewer. We applied a broadly used rubric based on seven types of misinformation: fabricated content, manipulated content, imposter content, false context, misleading content, false connection, and satire~\cite{Wardle2016}. We also added claims that could not be verified (inconclusive). Satire was not excluded because fake-news sites often label their content as satirical, and viral satire is often mistaken for real news. Further details about the verification procedure can be found in a technical report~\cite{2017arXiv170707592S}. Fig.~\ref{fig:piechart} shows that only a minority of claims in the collection (27\%) can be verified. The sampling method biases the analysis toward more prolific sources, some of which simply copy and past large numbers of articles from other sources. The fraction of verified claims is cut in half when sampling claims by tweets, thus biasing the sample toward popular rather than prolific sources~\cite{2017arXiv170707592S}.      

We also tracked the websites of several independent fact-checking organizations: \url{politifact.com}, \url{snopes.com}, \url{factcheck.org}, \url{badsatiretoday.com}, \url{hoax-slayer.com}, \url{opensecrets.org}, and \url{truthorfiction.com}. In April 2017 we added \url{climatefeedback.org}, which does not affect the present analysis.

\subsection{Hoaxy system architecture}

\begin{figure}
    \centering
    \includegraphics[width=\columnwidth]{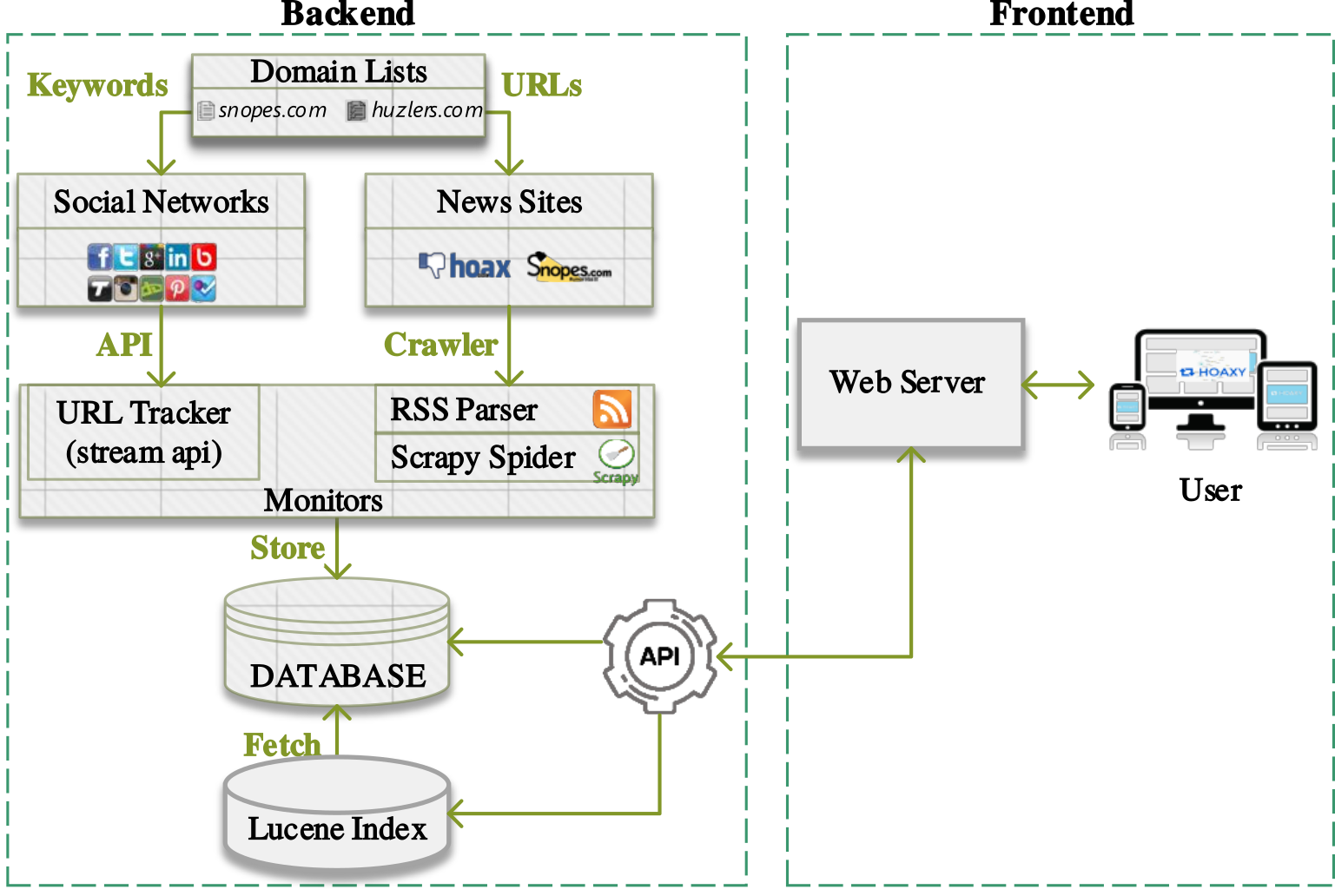}
    \caption{Hoaxy system architecture.}
    \label{fig:sys-arch}
\end{figure}

Fig.~\ref{fig:sys-arch} shows the architecture of the Hoaxy system. The system is composed of a back-end and a front-end. Next we describe some of the technical  aspects that went into the design and implementation of these components. 

\subsubsection{Back-end}

The back-end provides data collection, processing, storage, and indexing capabilities. We start from the list of sources discussed earlier. Data are collected from two realms: social media (i.e., Twitter) and the news source sites in the list. To collect data from Twitter, Hoaxy filters the real-time stream for tweets matching our list of domain keywords~\cite{TwitterStreamAPI}. Matches are performed server-side against the complete text of the tweet. This means that for each delivered tweet we further make sure that the match is actually a hyperlink. Tweets that simply mention our sources but do not link to them (e.g., ``I read this on snopes.com!'') are discarded.

All matching link URLs are then extracted from the tweet and fetched directly from the source. To get a complete snapshot of all content produced by the sources, Hoaxy also regularly crawls their websites in a separate process. We use a mix of RSS and direct crawling to do so. Regardless of the way it is collected, from each fetched document Hoaxy extracts title, metadata, and body information. 

All collected data (tweets and fetched documents) are saved in a relational database. Documents are further indexed using Lucene~\cite{lucene}, to enable full-text search from the front-end.    

Content duplication and document text extraction are two critical aspects of this data collection pipeline. Because we are crawling data from the Web, we expect to observe several different variants of the same URLs. This is especially true for the resources obtained from the social media stream, for which duplication may occur due to marketing campaign and other tracking parameters, shortening (e.g.,~\url{bit.ly}) and snapshotting (e.g.,~\url{archive.is}) services, and domain aliasing (e.g.,~\url{dcgazette.com} and~\url{thedcgazette.com}).  

While acknowledging that principled solutions to deal with the problem of Web content duplication have been around for decades~\cite{BRODER19971157}, we found that a set of few, simple heuristics gave satisfactory results. For example, we found that focusing on the most common tracking parameters (i.e. UTM parameters) we can canonicalize about 30\% of all URLs. Similarly, by following all types of HTTP redirect responses, we resolve  shortened URLs for about 45\% of the URLs extracted from tweets. Snapshotting and domain aliases account instead for only a handful of duplicates, and we simply ignore them.   

We also had the problem of extracting the actual text of the fetched documents. There is a lot of extraneous content in the body of documents due to the presence of ads, comment threads, and personalization. All this `noise' poses a problem for indexing the corpus efficiently. Algorithms for document text extraction have been around for several years~\cite{Gupta:2003:DCE:775152.775182}. We tested several implementations and eventually settled for the one offered by a third-party API~\cite{Mercury}. 

Having collected, processed, stored, and indexed all the data, the final component of the back-end is the API, a small piece of middleware that enables programmatic access to both the relational database and the full-text Lucene index for the purposes of search and visualization. 

\subsubsection{Front-end}

\begin{figure}
    \center
    \includegraphics[width=\columnwidth]{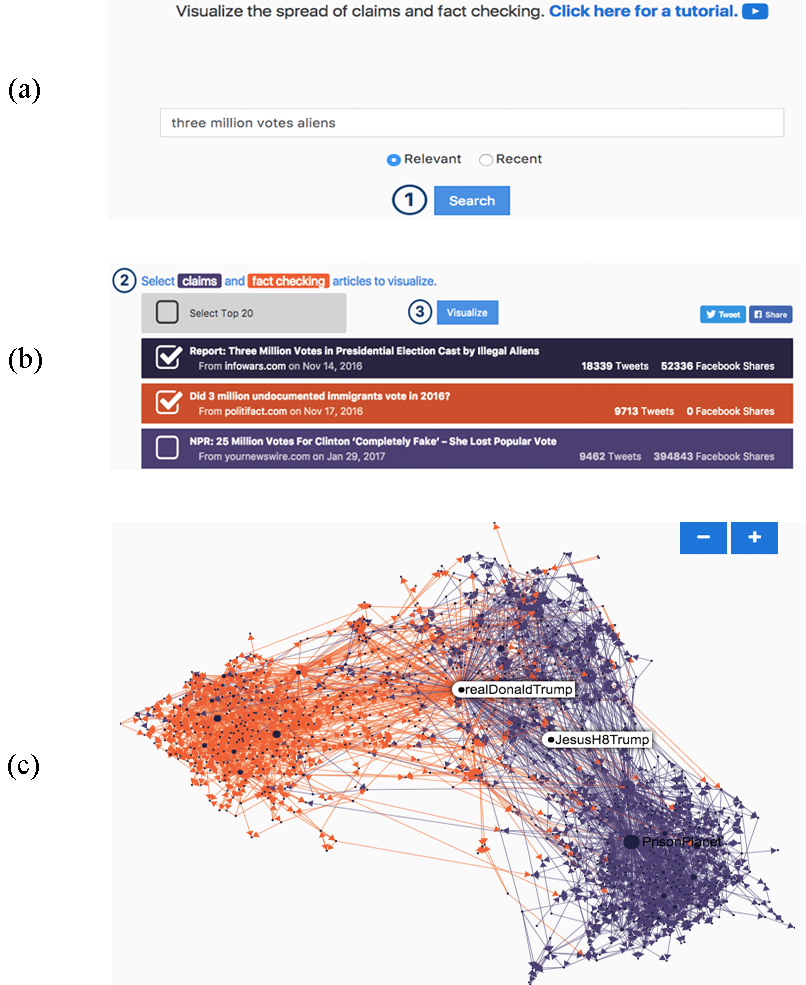}
    \caption{Screen shots from the user interface of Hoaxy: (a)~the user enters a query in the search engine interface; (b)~from the list of results, the user selects articles including claims (purple) and/or related fact-checking (orange) to visualize (colors online); (c)~a detail from the interactive network diffusion visualization for the query ``three million votes aliens.'' Edge colors represent the type of information exchanged. The network shown here displays strong polarization between claims and fact-checking, which is typical.}
    \label{fig:frontend}
\end{figure}

Hoaxy provides an intuitive Web interface to search and visualize the spread of claims contained in our database of misinformation, and the competition with subsequent fact-checking verifications (see Fig.~\ref{fig:frontend}). The user first specifies a query (Fig.~\ref{fig:frontend}(a)). Users can choose to retrieve either the most relevant or the most recent results. To do so, we first send the query to Lucene, which returns a list of most relevant/recent claims and fact-checking articles. In practice, because there are many more claim articles than fact-checking ones, and claims tend to outperform fact-checking in terms of popularity, we rank claims separately from fact-checks, and then merge the top results from the two rankings into a single list. Finally we re-rank the results based on the number of tweets in the database. 

After selecting the results that match their query (Fig.~\ref{fig:frontend}(b)), the user can finally visualize the results. Hoaxy provides two types of visualization: a timeline plot (not shown in the figure) that displays the growth in the number of tweets for both claims and fact-checking, and an interactive visualization of the diffusion network (Fig.~\ref{fig:frontend}(c)). In the  network, nodes represent Twitter accounts and edges connect any two users that exchanged information by retweet, reply, mention, or quoted retweet. Edge directionality represents the flow of information, e.g., from the retweeted to the retweeter account or from the mentioning to the mentioned account.

\subsubsection{Deployment}

We started collecting data with Hoaxy from 76 sources --- 69 of claims and 7 of fact-checking --- in June 2016. In December 2016, 50 more sources of claims were added.  
The system has collected data continuously ever since.
As of October 2017, Hoaxy has collected a total of 29,351,187 tweets --- 27,648,423 with links to claim sources and 1,705,576 with links to fact-checking sources. The total number of documents collected so far is 653,911 --- 628,350 by claim sources and 25,561 by fact-checking ones. 

\begin{figure}
    \centering
    \includegraphics[width=\columnwidth]{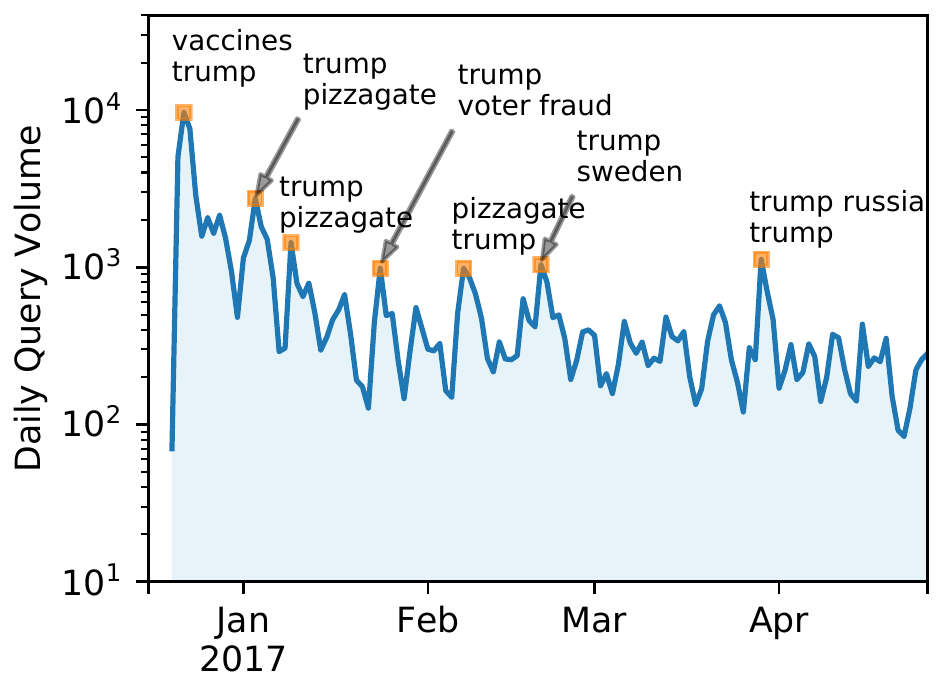}
    \caption{Usage of Hoaxy in terms of daily volume of queries since the launch of the public Web tool in December 2016. The two most frequent search terms are shown in correspondence to some of the main peaks of user activity.}
    \label{fig:hoaxy-usage}
\end{figure}

The public Web interface of Hoaxy was launched on December 20, 2016. Fig.~\ref{fig:hoaxy-usage} plots the daily query volume and some of the most popular topics queried by users over the course of the first 6 months of operation. Unsurprisingly, the term `Trump' is among the most popular search terms, but we also see substantial churn in user interest, with topics following closely the most popular pieces of controversial information of the moment, e.g., `vaccines,' `pizzagate,' `voter fraud' and `Trump Russia.'

\subsection{Datasets}

To explore our research questions, we focus on the retweet network (including quoted retweets) for links to either claims or fact-checking articles. A retweet provides information about the primary spreader (retweeted account) and secondary spreader (retweeting account). 
To be sure, Hoaxy does collect any kind of tweet, as long as a URL, whose Web domain matches our list of sources, is included in the tweet. To give an idea of the full scope of the Hoaxy dataset, retweets and quoted retweets occur 66.9\% of the times; approximately 1 in 10 retweets is a quoted retweet. Of the remaining types of tweets, replies (i.e. tweets forming a conversation thread and including an @-mention of another users) account for 2.1\% of the total. The remaining tweets are neither retweets nor replies; they are \emph{original} tweets. 

The network is a graph defined as follows: we include a node for each Twitter user account in the database. Edges are directed (as explained earlier) and weighted. The weight of an edge represents the number of retweets from one account to another. That is, we increase by one the weight on a directed edge from user $a$ to user $b$ every time we observe that $b$ retweets $a$. Edges are labeled by the type of content being retweeted. To do so, we split the total weight $w(e)$ of edge $e$ in two separate counts, one that keeps track of retweets of claims ($w_c$) and one of fact-checks ($w_f$), respectively. That is, $w(e) = w_c(e) + w_f(e)$ for all $e \in E$. We observe $w_c(e) \cdot w_f(e) > 0$ in only a small minority of edges, meaning that we can easily label each edge as a `claim' or `fact-check' edge with a simple majority rule (ties are broken at random). 

Because prior work shows that collective attention patterns change dramatically in the wake of highly anticipated events, like elections~\cite{Lehmann:2012:DCC:2187836.2187871}, we split our analysis in two periods, pre- and post-Election Day (Nov. 8, 2016). 
Table~\ref{tab:data} provides a summary of the three networks analyzed in this paper. We explore the overall spread of content on the full network spanning six months before Election Day, including both claims and fact-checking. We decompose this network into its $k$-core shells to uncover the functional roles of the most densely connected sub-graph. Row 1 of Table~\ref{tab:data} shows summary statistics for the network used at this stage.

\begin{table}
\begin{adjustwidth}{-1in}{0in}
  \caption{Summary of the data used in the network analysis. 
  $E_{f}$ is the set of edges labeled as `fact-check.'}
  \begin{minipage}{\linewidth}
  \renewcommand{\thefootnote}{\textit\alph{footnote}}
  \renewcommand\footnoterule{}
  \begin{tabular*}{0.98\linewidth}{@{\extracolsep{\fill}}lllrrr}
  \toprule
  & Network & Period & $|V|$ & $|E|$ & $|E_{f}|$ \\
  \midrule
  %
  1 & Claims + fact-checks & pre-election\footnotemark[1] & $346,573$ & $1,091,552$ & $279,283$ \\
  %
  2 & Claims only &
  pre-\footnotemark[1] + post-election\footnotemark[2] & 
  $630,368$ & $2,236,041$ & $0$ \\
  %
  3 & Claims only & pre-election\footnotemark[1] & 
  $227,363$ & $816,453$ & $0$  \\
  \bottomrule
  \end{tabular*}
  \footnotetext[1]{May 16, 2016 -- Nov. 7, 2016} \footnotetext[2]{Nov. 8, 2016 (Election Day) -- Oct. 9, 2017}
  \end{minipage}
  \label{tab:data}
\end{adjustwidth}
\end{table}

The second dataset is used to study the diffusion of the sole misinformation, therefore we ignore all edges labeled as `fact-check.' To characterize the long-term evolution of the main core, we extend the period of observation to October 9, 2017. Recall that 50 additional sources of claims were added to Hoaxy in December 2016. To keep our analysis consistent across the pre- and post-Election Day periods, we do not include data from these sites in the present work. The network in row 2 of Table~\ref{tab:data} is considered at this stage. 

The third dataset (row 3 of Table~\ref{tab:data}) includes only claims but goes back to the pre-Election-Day period to characterize the most central users in the core and the robustness of the network.

\section{Results}

Having described in the prior section how Hoaxy collects data, let us now analyze the misinformation diffusion networks. 
To the best of our knowledge, the following is the first in-depth analysis of the diffusion network of online misinformation and fact-checking in the period of the 2016 US Presidential Election.

\subsection{Claims vs.~fact-checking}

\begin{figure}
    \centering
    \includegraphics[width=0.95\columnwidth]{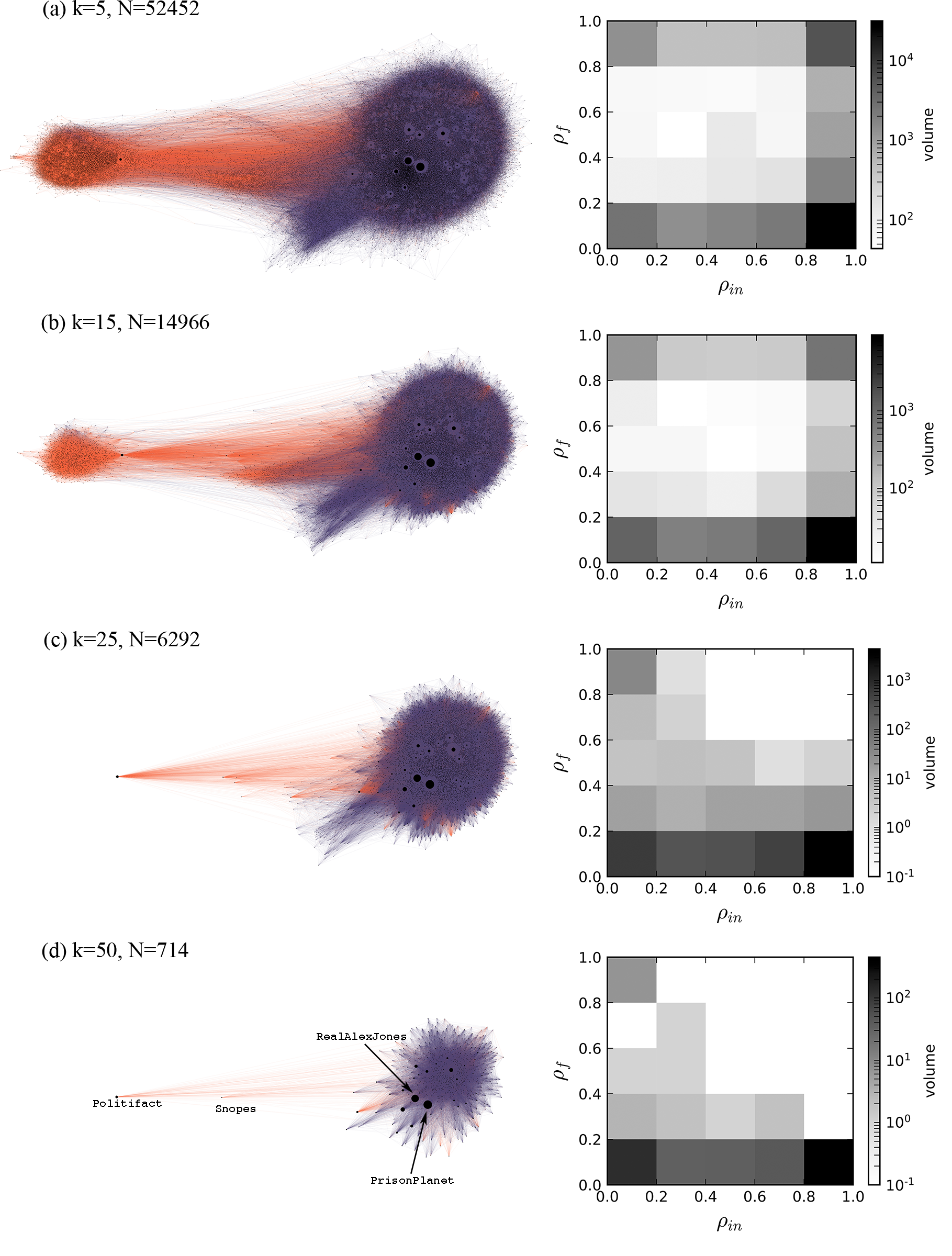}
    \caption{$k$-Core decomposition of the pre-Election retweet network collected by Hoaxy. Panels (a)-(d) show four different cores for values of $k=5,15,25,50$ respectively. Networks are visualized using a force-directed layout. Edge colors represent the type of content: orange for fact-checks and purple for claims (colors online). The innermost sub-graph (d), where each node has degree $k \ge 50$, corresponds to the main core. The heat maps show, for each core, the distribution of accounts in the space represented by two coordinates: the retweet ratio $\rho_{in}$ and the fact-checking ratio $\rho_{f}$ (see text).}
    \label{fig:kcore}
\end{figure}

We performed $k$-core decomposition of the entire network (row 1 of Table~\ref{tab:data}). Fig.~\ref{fig:kcore} visualizes different $k$-cores for increasing values of $k$. The main core is obtained at $\max\{k\} = 50$. We can draw several insights from these visualizations. First, the force-directed layout algorithm splits the network in two communities. There is substantial content segregation across these two communities, which we denote as the `fact-checkers' and (misinformation) `spreaders,' respectively. The edges across the two groups appear to be mostly colored in orange, suggesting some exposure of misinformation spreaders to fact-checks. The group of fact-checkers disappears as $k$ is gradually increased, moving toward the innermost, densest portion of the network.

However, it is still possible to appreciate \emph{some} retweeting of fact-checking content involving spreaders even in the main core (Fig.~\ref{fig:kcore}(d)).
To understand in more quantitative terms the role of fact-checking in the spread of information in the core, we characterize users according to two simple metrics. Recall that in a weighted, undirected network the \emph{strength} of a node is the sum of all the weights of all its incident edges, $s(v) = \sum_{e \in v} w(e)$. In a directed network one can likewise define the \emph{in-strength} $s_{in}$ and the \emph{out-strength} $s_{out}$, by taking the sum only on the incoming and outgoing edges, respectively. We further consider edge labels and distinguish between claim ($s_{\rm c}$) and fact-check ($s_f$) strength. For each node $v \in V$ let us define two ratios, the \emph{fact-checking ratio} $\rho_f$ and the \emph{retweet ratio} $\rho_{in}$:
\begin{eqnarray}
    \rho_f(v) &=& \frac{s_f(v)}{s_f(v) + s_c(v)} = \frac{s_f(v)}{s(v)} \\
    \rho_{in}(v) &=& \frac{s_{in}(v)}{s_{in}(v) + s_{out}(v)} = \frac{s_{in}(v)}{s(v)}.
\end{eqnarray}
Intuitively, a user with a value of $\rho_f$ close to unity is going to be a fact-checker (as opposed to claim spreader), whereas an account with a value of $\rho_{in}$ close to unity is going to be a secondary spreader of information, i.e., to amplify messages through retweets rather than post original messages. 
The right-hand side of Fig.~\ref{fig:kcore} shows the joint distributions of $(\rho_f, \rho_{in})$ for different values of $k$. We observe that for small values of $k$, most users fall close to the four corners of the space, meaning that they take on exactly one of the four possible combinations of roles (`primary claim spreader', `secondary claim spreader', etc.). 

\begin{figure}
    \centering
    \includegraphics[width=\columnwidth]{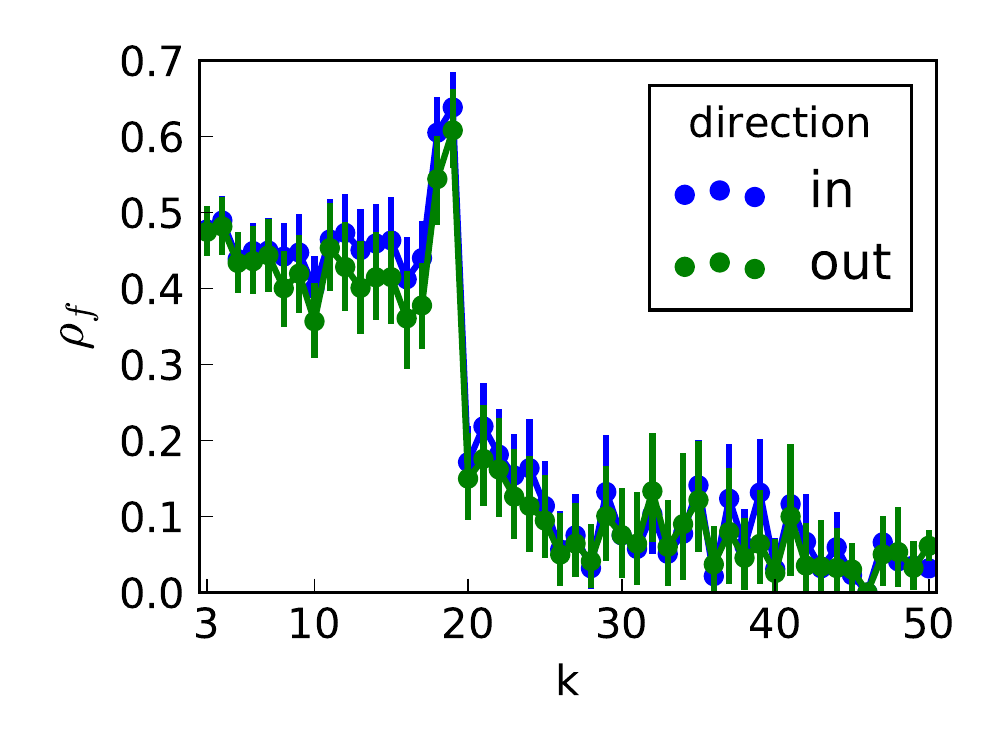}
    \caption{Average fact-checking ratio as a function of the shell number $k$ for activities of both primary spreading (`out') and secondary spreading (`in'). Error bars represent standard error.}
    \label{fig:factcheckactivity}
\end{figure}

For larger values of $k$ (Fig.~\ref{fig:kcore}(c,d)), we observe a shift away from secondary spreaders of fact-checking. In other words, the fact-checking links in the network core are retweeted by accounts who mainly spread misinformation. Fig.~\ref{fig:factcheckactivity} confirms that the drop in the spread of fact-checking is precipitous. For $k>20$ there is only a small, stable residual activity.

The fact that fact-checking still spreads in the main core is a somewhat surprising observation. Therefore we search for patterns that explain how claim spreaders interact with fact-checking. Manual inspection of the data let us identify three key characteristics of these retweets of fact-checking content made by spreaders in the main core: (1)~they link to fact-checking articles with biased, misleading wording; (2)~they attack fact-checking sites; or (3)~they use language that is inconsistent with the stance of a fact-checking article, for example implying that a claim is true even though the linked fact-checking article states that it is false. A sample of tweets with each of the aforementioned characteristics is shown in Table~\ref{tab:fcretweettable}. Similar patterns of citing mainstream media to challenge them have been observed by Starbird~\cite{starbird2017examining}.

\begin{table}
    \caption{Sample of tweets with fact-checking content published by accounts in the main core of the misinformation network.
    }
    \footnotesize
    \begin{tabular*}{\linewidth}{@{\extracolsep{\fill}}p{\linewidth}}
    \toprule
    \multicolumn{1}{c}{\sc Biased repetition} \\
    \midrule
    BREAKING NEWS! RINO GOP \#NeverTrump Leader PAID \$294K RINO TRAITORS prefer \#Hillary; Marxist SCOTUS Click \url{https://www.opensecrets.org/politicians/contrib.php?cid=N00035544\&cycle=2016\&type=I}\\
    \addlinespace
    HRC PRAISED HER KKK FRIEND MENTOR, BYRD! HRC IS RACIST! Hillary Kissed by Former Klan Member
    \url{http://www.snopes.com/clinton-byrd-photo-klan/}\\
    \addlinespace\midrule
    \multicolumn{1}{c}{\sc Attacks on fact-checking}\\
    \midrule
    Newsflash: Snopes itself is a biased left-wing Clinton mouthpiece  She knew he fooled the polygraph, was guilty \url{http://www.snopes.com/hillary-clinton-freed-child-rapist-laughed-about-it/}\\[1em]
    \addlinespace
    Lying Politifact caught telling another objective lie. CNN Is Hitler. \url{http://www.politifact.com/truth-o-meter/statements/2016/aug/23/donald-trump/donald-trump-fundraising-email-takes-cnn-anchors-c/}\\
    \addlinespace\midrule
    \multicolumn{1}{c}{\sc Inconsistency with fact-checking stance}\\
    \midrule
    13 Hours of HELL in Benghazi! No HELP was Sent?? Her E-Mails Show SHE KNEW THE TRUTH! \#LIAR \url{http://www.politifact.com/truth-o-meter/article/2016/feb/09/what-did-hillary-clinton-tell-families-people-who-/}\\
    \addlinespace
    Machado had sex on camera while filing a reality show. The media is lying about there being no sex tape. \url{http://www.snopes.com/alicia-machado-adult-star/}\\
    \bottomrule
    \end{tabular*}
    \label{tab:fcretweettable}
\end{table}

\subsection{Anatomy of the main core}

\subsubsection{Core dynamics}

Although we observe that the main core is dominated by misinformation spreaders, it is unclear if this has always been the case. From this point and in the subsequent analysis we discard all edges labeled as `fact-check' and focus only on the spread of misinformation. We start by investigating the long-term growth of the network. To do so we consider a network based on the Hoaxy dataset that extends post-Election Day; see row 2 of Table~\ref{tab:data}.  

\begin{figure}
    \centering
    \includegraphics[width=\columnwidth]{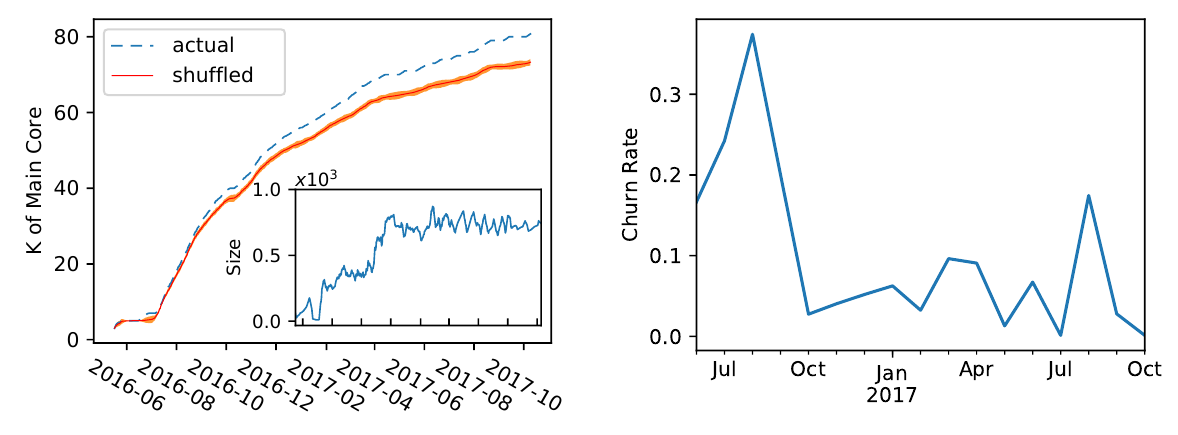}
    \caption{\emph{Left:}~Change of main core size and $k$ with the evolution of the network. A rolling window of one week is applied to filter fluctuations. The shuffled version is obtained by sampling from the configuration model. This is repeated many times to obtain the 95\% confidence interval shown in orange. The inset shows the size of the main core over time. \emph{Right:}~Churn rate (relative monthly change) of accounts in the main core.}
    \label{fig:network-growing}
\end{figure}

In particular we consider all retweets in our dataset in chronological order. At any given point in time, we consider a snapshot of the network formed by all retweets up to that point. We perform $k$-core decomposition on this cumulative network. We extract two pieces of information: the maximum value of $k$ and the size of the main core. The left panel of Fig.~\ref{fig:network-growing} shows how these two quantities change over time. We observe that the core gets both larger and denser. To characterize the extent to which the increasing density is just a byproduct of considering snapshots of growing size, we also plot the $k$ of the main core for a shuffled version of each snapshot (i.e., the configuration model of each snapshot). In rewiring edges, we preserve the degree sequence of the network.
While both the actual and the rewired network grow denser, the actual network does so at a higher rate, and the difference is statistically significant. 

\begin{figure}
    \centering
    \includegraphics[width=\columnwidth]{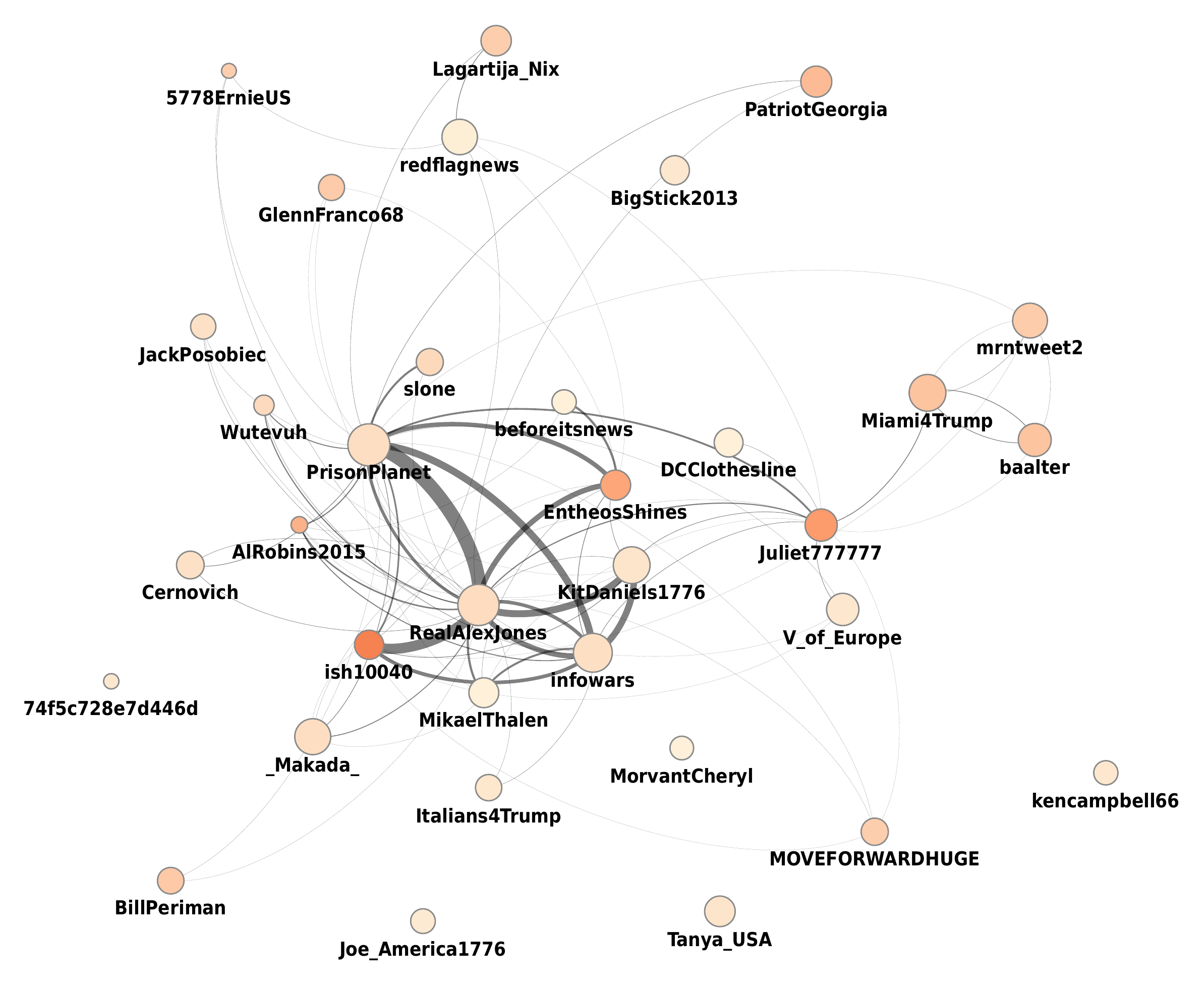}
    \caption{Retweet network of the stable main core of claim spreaders. Filtering by in-degree was applied to focus on the 34 accounts that retweet the most other accounts in the core. Node size represents out-degree (number of retweeters) and node color represents in-degree (color online).}
    \label{fig:stable-core}
\end{figure}

The main core reaches an equilibrium size of approximately 800 accounts around Election Day (see inset of left panel of Fig.~\ref{fig:network-growing}). This observations prompts the question of who are the users in the main core, and whether there is substantial churn in this group over time. By considering the intersection between two consecutive main cores in the sequence, we find a low churn rate after a peak in August (right panel of Fig.~\ref{fig:network-growing}), implying a stable set of users who consistently drive network activities. In fact, there are 321 users who remain in the main core for the whole duration of the observation period. The retweet network of a subset of them is shown in Fig.~\ref{fig:stable-core}. 

\subsubsection{Core membership}

We now return to the period before the election. For this  analysis we consider the pre-Election network, whose statistics are described in row 3 of Table~\ref{tab:data}. We use this network to characterize several aspects of how misinformation spreads before the election, starting with the presence of automated accounts. We do expect to find evidence of automation, given recent findings that show that a sizable fraction of the broader Twitter conversation about the elections was due to bots~\cite{FM7090}. An open question is whether bots were successful at spreading misinformation, which would be reflected by their tendency to locate in the core of the network, as opposed to the periphery. 

\begin{figure}
    \centering
     \includegraphics[width=\columnwidth]{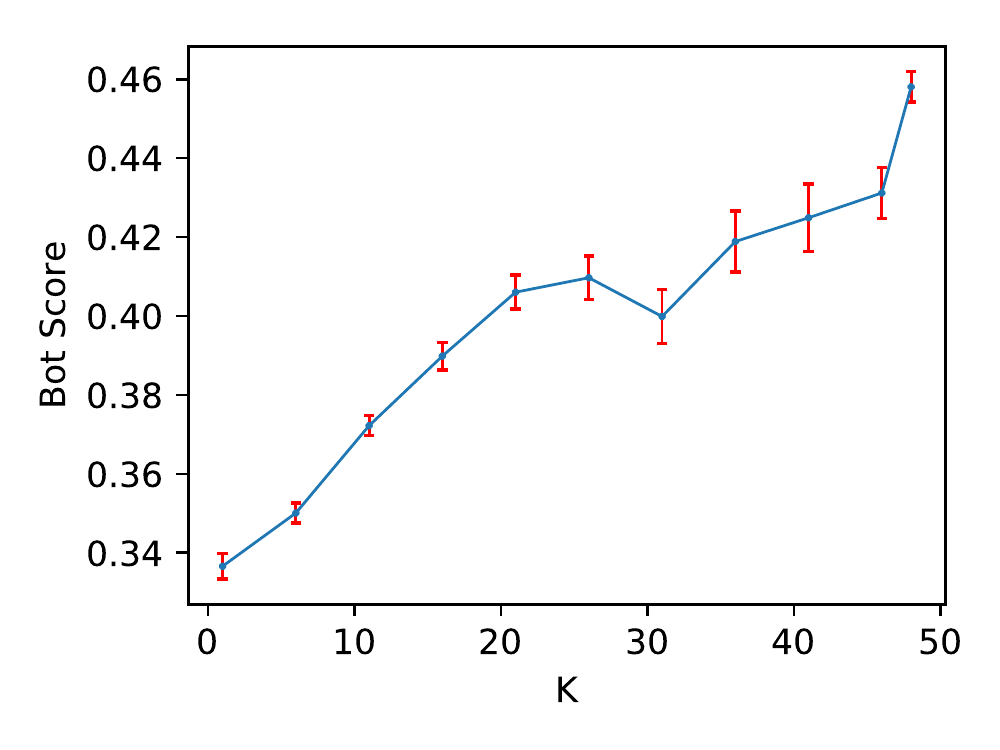}
    \caption{Average bot score for a random sample of accounts drawn from different $k$-shells of the pre-Election Day retweet network, as a function of $k$. Only retweets including links to sources of misinformation are considered. Error bars represent standard errors.}
    \label{fig:bot-by-kshell}
\end{figure}

To determine the likelihood that the observed patterns of activity in the core of the network are the result of the deployment of social bots, we perform bot detection on a sample of accounts. After $k$-core decomposition, for each $k$-shell, we sample 2000 accounts at random. If the size of a shell is smaller, we include the whole shell. To estimate the likelihood that each of these account is automated we compute a bot score by querying Botometer, a state-of-the-art bot detection tool (see Methods). Fig.~\ref{fig:bot-by-kshell} shows a sharp increase in bot score as we move toward the core of the network, confirming our hypothesis.

\begin{figure}
    \centering
    \includegraphics[width=\columnwidth]{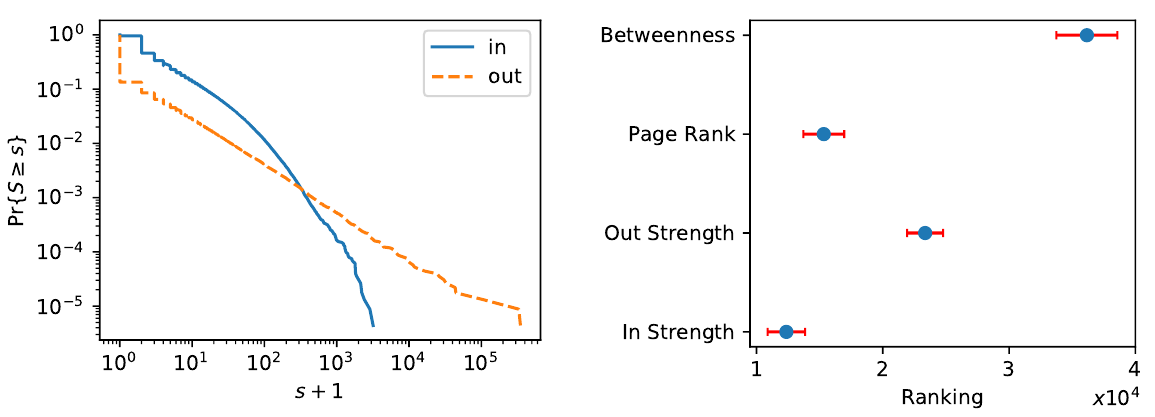}
    \caption{\emph{Left:}~Distribution of $s_{in}$ and $s_{out}$.
    \emph{Right:}~The average rank of users in the main core according to each centrality metric. Error bars represent standard errors.}
    \label{fig:claim-strength+ranking-mcore}
\end{figure}

For at least some of the most important individuals in the network core, it would be useful to have a behavioral description at a more granular level than just group averages. To this end, we need first to define a subset of important accounts. The network science toolbox provides us with several methods to identify salient nodes. We consider four centrality metrics: in-strength $s_{in}$, out-strength $s_{out}$, PageRank~\cite{page1999pagerank}, and betweenness~\cite{freeman1977set}. 
The first two are local measures of primary ($s_{out}$) and secondary ($s_{in}$) spreading activity. Intuitively, $s_{out}$ captures influence, as measured by the number of times that an account is retweeted. On the other hand, $s_{in}$ captures prolific accounts that retweet a lot. The distribution of $s_{in}$ and $s_{out}$ is shown in Fig.~\ref{fig:claim-strength+ranking-mcore} (left panel). Both distributions are broad, and the range of out-strength is broader than that of in-strength, due to the simple fact that, in networks of this size, the rate at which one can be retweeted is generally larger than that at which one can retweet others. The third and fourth measures are instead global notions of centrality, based on random walks and shortest paths, respectively. Given that these metrics capture fundamentally different notions of centrality, we expect them to produce different rankings of the nodes in the network. The right panel of Fig.~\ref{fig:claim-strength+ranking-mcore} shows strong variation in the average rank of users in the main core, confirming this intuition. The metric that seems to best capture the main core is the in-strength, indicating that a majority of core accounts are secondary spreaders (prolific accounts). 

For each measure, we rank the accounts in the main core and consider the top ten users. This exercise yields four lists of accounts, whose screen names are shown in Table~\ref{tab:central-spreaders}. There is little or no overlap between these lists, and their union yields 34 unique accounts. Having identified a subset of moderate size that includes main core members of potential interest, we performed a qualitative analysis of these accounts. Three human raters were asked to inspect the Twitter profile of each user independently, and to provide answers to the following questions: 
\begin{enumerate}
    \item Bot or Human? 
    \item Partisanship? 
    \item Personal or organizational account?
   \item How often does it share claims? 	
\end{enumerate}

\begin{table}
    \centering
    \footnotesize
    \caption{The top ten central users, ranked in descending order of centrality, in the claim network before the 2016 Election. Rankings are based on four different centrality metrics.}
    \begin{tabular*}{\columnwidth}{@{\extracolsep{\fill}}llll}
    \toprule
    Betweenness & PageRank & $s_{in}$ &	$s_{out}$ \\
    \midrule
    PrisonPlanet & ImmoralReport & PATROIT73 & RealAlexJones\\ 
    AlternativViewz & BillPeriman & BadCompany709 & PrisonPlanet\\ 
    RealAlexJones & alllibertynews & LovToRideMyTrek & infowars\\ 
    libertytarian & eavesdropann & PhilDeCarolis & redflagnews \\ 
    eavesdropann & Lagartija\_Nix & Roostrwoodstock & Miami4Trump\\
    BillPeriman & MMUSA\_PPN & Skinner147 & beforeitsnews\\ 
    wvufanagent99 & retireleo & MrNoahItALL & KitDaniels1776\\ 
    Miami4Trump & Nuevomedio & RESPECTPUNK434 & V\_of\_Europe\\ 
    Juliet777777 & ish10040 & Rubenz1133 & \_Makada\_\\ 
    ish10040 & EntheosShines & Cecil3695Cecil & Tanya\_USA\\
    \bottomrule
    \end{tabular*}
    \label{tab:central-spreaders}
\end{table}

For questions 1--3, whose answer is a categorical variable, the raters could also choose `neither' or `not sure'. After the annotators coded each account, for each question we applied a majority rule to identify the consensus label. The few cases in which a consensus could not be reached were broken by a fourth rater (one of the authors). The results are shown in Table~\ref{tab:labeling}. We report results for 32 of the 34 original accounts, since two accounts had been suspended by Twitter, and thus could not be annotated. Many of the central accounts appear to be automated and display a high degree of partisanship, all in support of the same candidate. 

\begin{table}
    \caption{Annotation of central users. For categorical questions (1--3), the top most frequent label, and its frequency, are reported. The question about claim sharing frequency (4) was on a 5-point Likert scale; we report the mean and standard deviation
    of the answers.}
    \centering
    \footnotesize
    \begin{tabular*}{\columnwidth}{@{\extracolsep{\fill}}rcccc}
    \toprule
    & \textbf{Betweenness} & \textbf{PageRank} & \textbf{$s_{in}$} & \textbf{$s_{out}$} \\
    \cmidrule{2-5}
    \multicolumn{5}{c}{\sc Bot/Human}\\
    \cmidrule{2-5}
    Top  & Bot & Bot & Bot & Bot \\
    Freq. & 6 & 6 & 4 & 4 \\
    \cmidrule{2-5}
    \multicolumn{5}{c}{\sc Partisanship}\\
    \cmidrule{2-5}
    Top  & Partisan & Partisan & Partisan & Partisan \\
    Freq. & 10 & 7 & 7 & 8 \\
    \cmidrule{2-5}
    \multicolumn{5}{c}{\sc Personal/Organizational}\\
    \cmidrule{2-5}
    Top  & Personal & Personal & Personal & Personal \\
    Freq. & 9 & 5 & 8 & 6 \\
    \cmidrule{2-5}
    \multicolumn{5}{c}{\sc How often does it share claims?}\\
    \cmidrule{2-5}
    Mean & $2.9 \pm 0.8$ & $2.4 \pm 0.9$ & $2.6 \pm 0.7$ & $3 \pm 1$\\
    \bottomrule
    \end{tabular*}
    \label{tab:labeling}
\end{table}

\subsection{Network robustness}

\begin{figure}
    \centering
    \includegraphics[width=\columnwidth]{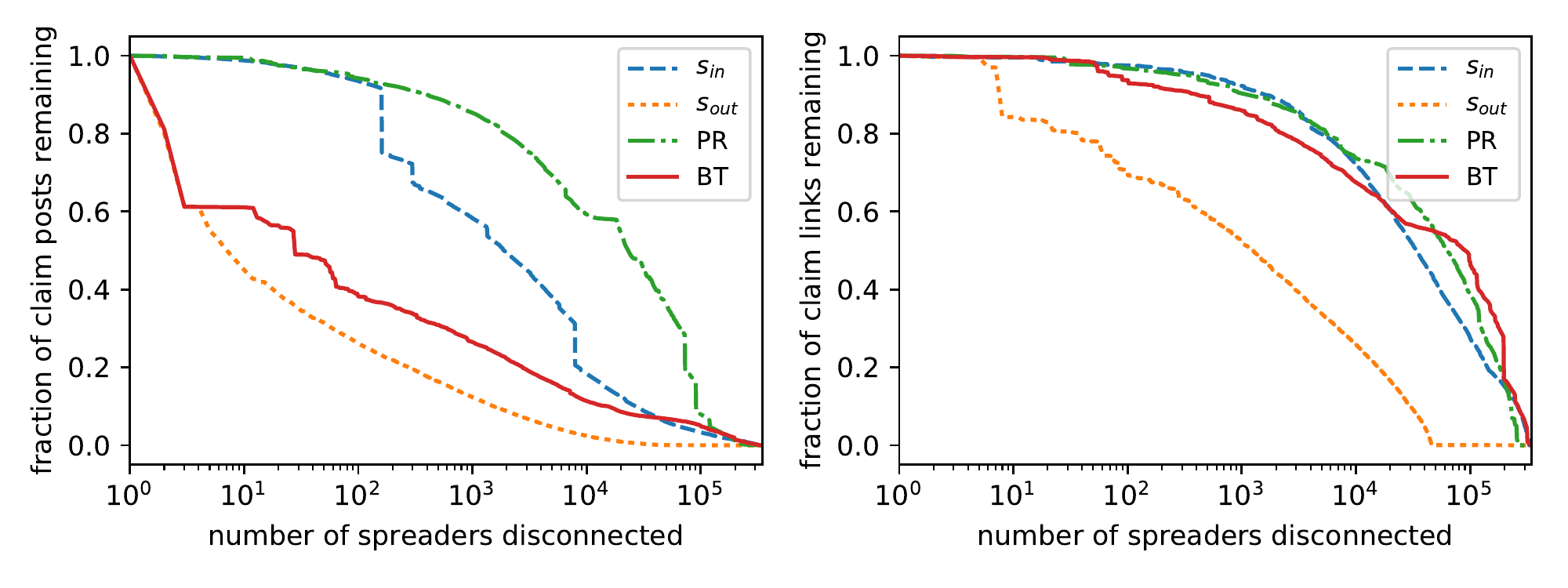}
    \caption{
    \emph{Left:}~Fraction of the retweets remaining vs.~number of spreaders disconnected in the network. 
    \emph{Right:}~Fraction of unique claim links remaining vs.~number of spreaders disconnected in the network. The priority of disconnected users is determined by ranking on the basis of different centrality metrics: $s_{in}$, $s_{out}$, betweenness (BT), and PageRank (PR).}
    \label{fig:claim-robust}
\end{figure}

Our last question is about the overall robustness of the network (row 3 of Table~\ref{tab:data}). We ask: \emph{How much does the efficient spread of claims rely on the activity of the most central nodes?} To explore this question we apply node disconnection, a standard procedure for estimating robustness in network science~\cite{PhysRevLett.85.5468}. The idea is to remove one node at a time, and analyze how two simple metrics are curtailed as a result: total volume of claim retweets, and total number of unique claim links. The more these quantities can be reduced by removing a small number of nodes, the more the efficiency of the misinformation network is disrupted.
We measure the fraction of retweets remaining after simulating the scenario in which a certain number of accounts are disconnected, by removing all edges to and from those accounts. We prioritize accounts to disconnect based on the four centrality metrics discussed before ($s_{in}$, $s_{out}$, betweenness, and PageRank). Fig.~\ref{fig:claim-robust} shows the result of the simulation. The greedy strategy that ranks users by decreasing out-strength achieves the best reduction of both metrics. The efficiency of the network is greatly impaired even after disconnecting as few as 10 most influential accounts (i.e., with greatest $s_{out}$). Surprisingly, disconnecting nodes with the highest $s_{in}$ is not an efficient strategy for reducing claims; the network is robust with respect to the removal of bots and other prolific accounts in the core. Betweenness, in comparison, seems to give good results on the total number of retweets (left panel of Fig.~\ref{fig:claim-robust}), but does not produce better results than PageRank and in-strength when considering unique links (right panel).

From a policy perspective, we are not proposing that a social media platform should suspend accounts whose posts are highly retweeted. Of course, platforms must take great care in minimizing the chances that a legitimate account is suspended. However, platforms do use various signals to identify and penalize low-quality information~\cite{FB-LOW-Q,Google-LOW-Q}. The present analysis suggests that the use of $s_{out}$ in the claim spreading network might provide a useful signal to prioritize further review, with the goal of  mitigating the spread of misinformation. Such an approach assumes the availability of a list of low-quality sources, which can be readily compiled.

\section{Discussion}

The rise of digital misinformation is calling into question the integrity of our information ecosystem. Here we made two contributions to the ongoing debate on how to best combat this threat.
First, we presented Hoaxy, an open platform that enables large-scale, systematic studies of how misinformation and fact-checking spread and compete on Twitter. We described key issues in its design and implementation. All Hoaxy data is available through an open API. Second, using data from Hoaxy, we presented an in-depth analysis of the misinformation diffusion network in the run up to and wake of the 2016 US Presidential Election. We found that the network is strongly segregated along the two types of information circulating in it, and that a dense, stable core emerged after the election. We characterized the main core in terms of multiple centrality measures and proposed an efficient strategies to reduce the circulation of information by penalizing key nodes in this network. The networks used in the present analysis are available on an institutional repository (see Methods).

Recall that Hoaxy collects 100\% of the tweets carrying each piece of misinformation in our collection, not a sample. As a result, our analysis provides a complete picture of the anatomy of the misinformation network. Of course, our methodology has some unavoidable limitations. First of all, Hoaxy only tracks a fixed, limited set of sources, due to data volume restrictions in the public Twitter API. Of these sources, it only tracks how their content spreads on Twitter, ignoring other social media platforms. Facebook, by far the largest social media platform, does not provide access to data on shares, ostensibly for privacy reasons, even though a significant fraction of misinformation spreads via its pages~\cite{del2016spreading}, which are understood to be public. Thus we acknowledge that coverage of our corpus of misinformation is incomplete. Nonetheless, by focusing on sources that have come to the attention of large media and fact-checking organizations, and that have been flagged as the most popular purveyors of unverified claims, Hoaxy captures a broad snapshot of  misinformation circulating online. 

Second, Hoaxy does not track the spread of unsubstantiated claims in the professional mainstream press. News websites do report unverified claims, in a manner and with a frequency dictated by their own editorial standards. For example, hedging language is often used to express degrees of uncertainty~\cite{Silverman2015}. While most claims reported in the mainstream media are eventually verified, many remain unverified, and some even turn out to be false. Some instances of misinformation may see their spread boosted as a result of  additional exposure on mainstream news outlets. Understanding the dynamics of the broader media and information ecosystem is therefore needed to fully comprehend the phenomenon of digital misinformation, but it is outside the scope of the present work.   

Third, we consider only US-based sources publishing English content. This is an unavoidable consequence of our reliance on lists produced by US-based media organizations. Different sources will be of course active in different countries. Worrisome amounts of misinformation, for example, have been observed in the run-up to the general elections in France~\cite{FM8005}. 
To foster the study of misinformation in non-US contexts, we have released the code of Hoaxy under an open-source license, so that other groups can build upon our work~\cite{HoaxyBackend,HoaxyFrontend}.

Last but not least, it is important to reiterate that the claims collected by Hoaxy are in general not verified. Inspection of our corpus confirms that not all claims collected by Hoaxy are completely inaccurate. As far as the present analysis is concerned, we provide an assessment of the rate of confirmed claims in our dataset (see Methods). When used as a search engine for misinformation, Hoaxy addresses this limitation by showing the most relevant fact-checking articles matching the input query, thereby facilitating claim verification. We hope that the data, software, and visualizations offered by the Hoaxy platform will be useful to researchers, reporters, policymakers, and, last but not least, ordinary Internet users as they learn to cope with online misinformation.

\paragraph{Acknowledgments.}

We are grateful to Ben Serrette and Valentin Pentchev of the Indiana University Network Science Institute (\url{iuni.iu.edu})
for supporting the development of the Hoaxy platform. Mihai Avram, Zackary Dunivin, Gregory Maus, and Vincent Wong annotated the most central users. Clayton A. Davis developed the Botometer API. Onur Varol and Nic Dias were instrumental in verification of the article sample. We are also indebted to Twitter for providing data through their API. C.S. thanks the Center for Complex Networks and Systems Research (\url{cnets.indiana.edu}) for the hospitality during his visit at the Indiana University School of Informatics and Computing. 
C.S. was supported by the China Scholarship Council. X.J. was supported in part by the National Natural Science Foundation of China (No. 61272010). G.L.C. was supported by IUNI. The development of the Botometer platform was supported in part by DARPA (grant W911NF-12-1-0037). A.F. and F.M. were supported in part by the James S. McDonnell Foundation (grant 220020274) and the National Science Foundation (award  CCF-1101743). The funders had no role in study design, data collection and analysis, decision to publish or preparation of the manuscript.

\bibliography{refs}

\begin{thebibliography}{10}

\bibitem{Pew2016}
M.~Barthel, A.~Mitchell, and J.~Holcomb, ``Many americans believe fake news is
  sowing confusion,'' Dec. 2016.

\bibitem{Pew2017a}
J.~Gottfried and E.~Shearer, ``News use across social media platforms 2017,''
  Sept. 2017.

\bibitem{Pew2017b}
M.~Barthel and A.~Mitchell, ``Americans' attitudes about the news media deeply
  divided along partisan lines,'' May 2017.

\bibitem{Ratkiewicz:2011:TMS:1963192.1963301}
J.~Ratkiewicz, M.~Conover, M.~Meiss, B.~Gon\c{c}alves, S.~Patil, A.~Flammini,
  and F.~Menczer, ``Truthy: Mapping the spread of astroturf in microblog
  streams,'' in {\em Proceedings of the 20th International Conference Companion
  on World Wide Web}, WWW '11, (New York, NY, USA), pp.~249--252, ACM, 2011.

\bibitem{W2015HiddenHands}
W.~Xiang, Z.~Zhilin, Y.~Xiang, J.~Yan, Z.~Bin, and L.~Shasha, ``Finding the
  hidden hands: a case study of detecting organized posters and promoters in
  sina weibo,'' {\em China Communications}, vol.~12, pp.~1--13, November 2015.

\bibitem{ICWSM112850}
J.~Ratkiewicz, M.~Conover, M.~Meiss, B.~Goncalves, A.~Flammini, and F.~Menczer,
  ``Detecting and tracking political abuse in social media,'' in {\em Proc.
  International AAAI Conference on Web and Social Media}, (Palo Alto, CA),
  pp.~297--304, AAAI, 2011.

\bibitem{Sampson:2016:LIS:2983323.2983697}
J.~Sampson, F.~Morstatter, L.~Wu, and H.~Liu, ``Leveraging the implicit
  structure within social media for emergent rumor detection,'' in {\em
  Proceedings of the 25th ACM International on Conference on Information and
  Knowledge Management}, CIKM '16, (New York, NY, USA), pp.~2377--2382, ACM,
  2016.

\bibitem{Wu2016}
L.~Wu, F.~Morstatter, X.~Hu, and H.~Liu, ``Mining misinformation in social
  media,'' in {\em Big Data in Complex and Social Networks} (M.~T. Thai, W.~Wu,
  and H.~Xiong, eds.), Business \& Economics, pp.~125--152, Boca Raton, FL: CRC
  Press, Dec 2016.

\bibitem{Declerck2016}
T.~Declerck, P.~Osenova, G.~Georgiev, and P.~Lendvai, ``Ontological modelling
  of rumors,'' in {\em Linguistic Linked Open Data: 12th EUROLAN 2015 Summer
  School and RUMOUR 2015 Workshop, Sibiu, Romania, July 13-25, 2015, Revised
  Selected Papers} (D.~Trandab{\u{a}}{\c{T}} and D.~G{\^i}fu, eds.), pp.~3--17,
  Berlin/Heidelberg, Germany: Springer International Publishing, 2016.

\bibitem{Kumar:2016:DWI:2872427.2883085}
S.~Kumar, R.~West, and J.~Leskovec, ``Disinformation on the web: Impact,
  characteristics, and detection of wikipedia hoaxes,'' in {\em Proceedings of
  the 25th International Conference on World Wide Web}, WWW '16, (Republic and
  Canton of Geneva, Switzerland), pp.~591--602, International World Wide Web
  Conferences Steering Committee, 2016.

\bibitem{Varol2017}
O.~Varol, E.~Ferrara, F.~Menczer, and A.~Flammini, ``Early detection of
  promoted campaigns on social media,'' {\em EPJ Data Science}, vol.~6, p.~13,
  Jul 2017.

\bibitem{botornot_icwsm17}
O.~Varol, E.~Ferrara, C.~A. Davis, F.~Menczer, and A.~Flammini, ``Online
  human-bot interactions: Detection, estimation, and characterization,'' in
  {\em Proc. International AAAI Conference on Web and Social Media}, (Palo
  Alto, CA), pp.~280--289, AAAI, 2017.

\bibitem{Ferrara:2016:RSB:2963119.2818717}
E.~Ferrara, O.~Varol, C.~Davis, F.~Menczer, and A.~Flammini, ``The rise of
  social bots,'' {\em Commun. ACM}, vol.~59, pp.~96--104, June 2016.

\bibitem{FM8005}
E.~Ferrara, ``Disinformation and social bot operations in the run up to the
  2017 french presidential election,'' {\em First Monday}, vol.~22, no.~8,
  2017.

\bibitem{2017arXiv170707592S}
C.~Shao, G.~L. Ciampaglia, O.~Varol, A.~Flammini, and F.~Menczer, ``The spread
  of misinformation by social bots,'' e-print arXiv:1707.07592, CoRR, 2017.

\bibitem{ECKER2017185}
U.~K. Ecker, J.~L. Hogan, and S.~Lewandowsky, ``Reminders and repetition of
  misinformation: Helping or hindering its retraction?,'' {\em Journal of
  Applied Research in Memory and Cognition}, vol.~6, no.~2, pp.~185--192, 2017.

\bibitem{Nyhan2016}
B.~Nyhan and J.~Reifler, ``Estimating fact-checking's effects,'' Aug. 2016.

\bibitem{Jun06062017}
Y.~Jun, R.~Meng, and G.~V. Johar, ``Perceived social presence reduces
  fact-checking,'' {\em Proceedings of the National Academy of Sciences},
  vol.~114, no.~23, pp.~5976--5981, 2017.

\bibitem{2017arXiv170700574N}
A.~Nematzadeh, G.~L. Ciampaglia, F.~Menczer, and A.~Flammini, ``How algorithmic
  popularity bias hinders or promotes quality,'' e-print arXiv:1707.00574,
  CoRR, 2017.

\bibitem{Qiu2017}
X.~Qiu, D.~F.~M.~Oliveira, A.~Sahami~Shirazi, A.~Flammini, and F.~Menczer,
  ``Limited individual attention and online virality of low-quality
  information,'' {\em Nature Human Behavior}, vol.~1, pp.~0132--, June 2017.

\bibitem{Wardle2016}
C.~Wardle, ``{Fake news. It's complicated.},'' white paper, First Draft News,
  February 2017.

\bibitem{ICWSM1510582}
T.~Mitra and E.~Gilbert, ``Credbank: A large-scale social media corpus with
  associated credibility annotations,'' in {\em Proc. International AAAI
  Conference on Web and Social Media}, (Palo Alto, CA), pp.~258--267, AAAI,
  2015.

\bibitem{Hassan:2015:DCF:2806416.2806652}
N.~Hassan, C.~Li, and M.~Tremayne, ``Detecting check-worthy factual claims in
  presidential debates,'' in {\em Proceedings of the 24th ACM International on
  Conference on Information and Knowledge Management}, CIKM '15, (New York, NY,
  USA), pp.~1835--1838, ACM, 2015.

\bibitem{Metaxas:2015:UTI:2685553.2702691}
P.~T. Metaxas, S.~Finn, and E.~Mustafaraj, ``Using twittertrails.com to
  investigate rumor propagation,'' in {\em Proceedings of the 18th ACM
  Conference Companion on Computer Supported Cooperative Work \& Social
  Computing}, CSCW'15 Companion, (New York, NY, USA), pp.~69--72, ACM, 2015.

\bibitem{ICWSM1510592}
S.~Carton, S.~Park, N.~Zeffer, E.~Adar, Q.~Mei, and P.~Resnick, ``Audience
  analysis for competing memes in social media,'' in {\em Proc. International
  AAAI Conference on Web and Social Media}, (Palo Alto, CA), pp.~41--50, AAAI,
  2015.

\bibitem{Emergent.info}
C.~Silverman, ``Emergent,'' 2015.

\bibitem{Ciampaglia2017}
G.~L. Ciampaglia, A.~Mantzarlis, G.~Maus, and F.~Menczer, ``Research challenges
  of digital misinformation: Toward a trustworthy web,'' {\em AI Magazine},
  vol.~in press, 2018.

\bibitem{Lazer2017}
D.~Lazer, M.~Baum, N.~Grinberg, L.~Friedland, K.~Joseph, W.~Hobbs, and
  C.~Mattsson, ``Combating fake news: An agenda for research and action,'' Feb.
  2017.

\bibitem{Lu2014Network}
X.~Lu and C.~Brelsford, ``Network structure and community evolution on twitter:
  Human behavior change in response to the 2011 japanese earthquake and
  tsunami,'' {\em Scientific Reports}, vol.~4, p.~6773, 2014.

\bibitem{del2016spreading}
M.~Del~Vicario, A.~Bessi, F.~Zollo, F.~Petroni, A.~Scala, G.~Caldarelli, H.~E.
  Stanley, and W.~Quattrociocchi, ``The spreading of misinformation online,''
  {\em Proc. National Academy of Sciences}, vol.~113, no.~3, pp.~554--559,
  2016.

\bibitem{Schmidt21032017}
A.~L. Schmidt, F.~Zollo, M.~Del~Vicario, A.~Bessi, A.~Scala, G.~Caldarelli,
  H.~E. Stanley, and W.~Quattrociocchi, ``Anatomy of news consumption on
  facebook,'' {\em Proceedings of the National Academy of Sciences}, vol.~114,
  no.~12, pp.~3035--3039, 2017.

\bibitem{starbird2017examining}
K.~Starbird, ``Examining the alternative media ecosystem through the production
  of alternative narratives of mass shooting events on twitter.,'' in {\em
  Proceedings of the International AAAI Conference on Web and Social Media
  (ICWSM)}, pp.~230--239, 2017.

\bibitem{Silverman2016}
C.~Silverman, ``Viral fake election news outperformed real news on facebook in
  final months of the us election,'' Nov. 2016.

\bibitem{FB-INFO-OPS}
J.~Weedon, W.~Nuland, and A.~Stamos, ``Information operations and facebook,''
  Apr. 2017.

\bibitem{FB-LOW-Q}
A.~Mosseri, ``News feed fyi: Showing more informative links in news feed,''
  June 2017.

\bibitem{TwitterBlog}
C.~Crowell, ``Our approach to bots \& misinformation,'' June 2017.

\bibitem{shao2016hoaxy}
C.~Shao, G.~L. Ciampaglia, A.~Flammini, and F.~Menczer, ``Hoaxy: A platform for
  tracking online misinformation,'' in {\em Proceedings of the 25th
  International Conference Companion on World Wide Web}, WWW '16 Companion,
  (Republic and Canton of Geneva, Switzerland), pp.~745--750, International
  World Wide Web Conferences Steering Committee, 2016.

\bibitem{Brandtzaeg:2017:TDO:3134526.3122803}
P.~B. Brandtzaeg and A.~F{\o}lstad, ``Trust and distrust in online
  fact-checking services,'' {\em Commun. ACM}, vol.~60, pp.~65--71, Aug. 2017.

\bibitem{Google-LOW-Q}
B.~Gomes, ``Our latest quality improvements for search,'' Apr. 2017.

\bibitem{HoaxyAPIDoc}
C.~Shao, F.~Menczer, and G.~L. Ciampaglia, ``Hoaxy api documentation,'' Oct.
  2017.

\bibitem{PhysRevLett.96.040601}
S.~N. Dorogovtsev, A.~V. Goltsev, and J.~F.~F. Mendes, ``$k$-core organization
  of complex networks,'' {\em Phys. Rev. Lett.}, vol.~96, p.~040601, Feb 2006.

\bibitem{Alvarez-Hamelin2008}
J.~I. Alvarez-Hamelin, L.~Dall'Asta, A.~Barrat, and A.~Vespignani, ``K-core
  decomposition of internet graphs: hierarchies, self-similarity and
  measurement biases,'' {\em Networks and Heterogeneous Media}, vol.~3,
  pp.~371--393, June 2008.

\bibitem{Kitsak2010}
M.~Kitsak, L.~K. Gallos, S.~Havlin, F.~Liljeros, L.~Muchnik, H.~E. Stanley, and
  H.~A. Makse, ``Identification of influential spreaders in complex networks,''
  {\em Nature Physics}, vol.~6, pp.~888--, Aug. 2010.

\bibitem{conover12partisan}
M.~D. Conover, B.~Gon\c{c}alves, A.~Flammini, and F.~Menczer, ``Partisan
  asymmetries in online political activity,'' {\em EPJ Data Science}, vol.~1,
  p.~6, June 2012.

\bibitem{Davis16BotOrNot}
C.~A. Davis, O.~Varol, E.~Ferrara, A.~Flammini, and F.~Menczer, ``Botornot: A
  system to evaluate social bots,'' in {\em Proceedings of the 25th
  International Conference Companion on World Wide Web}, WWW '16 Companion,
  (Republic and Canton of Geneva, Switzerland), pp.~273--274, International
  World Wide Web Conferences Steering Committee, 2016.

\bibitem{BotometerAPI}
C.~A. Davis, ``Botometer api,'' Oct. 2017.

\bibitem{HoaxyFAQ}
C.~Shao, F.~Menczer, and G.~L. Ciampaglia, ``Hoaxy faq,'' 2017.

\bibitem{TwitterStreamAPI}
Twitter, ``Filter realtime tweets,'' Oct. 2017.

\bibitem{lucene}
{Apache Software Foundation}, ``Apache lucene,'' 2005.

\bibitem{BRODER19971157}
A.~Z. Broder, S.~C. Glassman, M.~S. Manasse, and G.~Zweig, ``Syntactic
  clustering of the web,'' {\em Computer Networks and ISDN Systems}, vol.~29,
  no.~8, pp.~1157--1166, 1997.
\newblock Papers from the Sixth International World Wide Web Conference.

\bibitem{Gupta:2003:DCE:775152.775182}
S.~Gupta, G.~Kaiser, D.~Neistadt, and P.~Grimm, ``Dom-based content extraction
  of html documents,'' in {\em Proceedings of the 12th International Conference
  on World Wide Web}, WWW '03, (New York, NY, USA), pp.~207--214, ACM, 2003.

\bibitem{Mercury}
P.~L. LLC, ``Mercury web parser by postlight,'' Oct. 2017.

\bibitem{Lehmann:2012:DCC:2187836.2187871}
J.~Lehmann, B.~Gon\c{c}alves, J.~J. Ramasco, and C.~Cattuto, ``Dynamical
  classes of collective attention in twitter,'' in {\em Proceedings of the 21st
  International Conference on World Wide Web}, WWW '12, (New York, NY, USA),
  pp.~251--260, ACM, 2012.

\bibitem{FM7090}
A.~Bessi and E.~Ferrara, ``Social bots distort the 2016 u.s. presidential
  election online discussion,'' {\em First Monday}, vol.~21, no.~11, 2016.

\bibitem{page1999pagerank}
L.~Page, S.~Brin, R.~Motwani, and T.~Winograd, ``The pagerank citation ranking:
  Bringing order to the web.,'' tech. rep., Stanford InfoLab, 1999.

\bibitem{freeman1977set}
L.~C. Freeman, ``A set of measures of centrality based on betweenness,'' {\em
  Sociometry}, vol.~40, no.~1, pp.~35--41, 1977.

\bibitem{PhysRevLett.85.5468}
D.~S. Callaway, M.~E.~J. Newman, S.~H. Strogatz, and D.~J. Watts, ``Network
  robustness and fragility: Percolation on random graphs,'' {\em Phys. Rev.
  Lett.}, vol.~85, pp.~5468--5471, Dec 2000.

\bibitem{Silverman2015}
C.~Silverman, ``Lies, damn lies and viral content: How news websites spread
  (and debunk) online rumors, unverified claims and misinformation,''
  tow/knight report, Tow Center for Digital Journalism, Feb. 2015.

\bibitem{HoaxyBackend}
C.~Shao, F.~Menczer, and G.~L. Ciampaglia, ``Hoaxy backend,'' 2017.

\bibitem{HoaxyFrontend}
C.~Shao, L.~Wang, B.~Serrette, V.~Pentchev, F.~Menczer, and G.~L. Ciampaglia,
  ``Hoaxy frontend,'' 2017.

\end{thebibliography}

\end{document}